\documentclass[twocolumn,preprintnumbers,amsmath,amssymb,showpacs,prb]{revtex4}

\pdfoutput=1 % for arXiv

\usepackage{graphicx}% Include figure files
\usepackage{bm}% bold math
\usepackage{dcolumn}
\usepackage{amssymb}
\usepackage{amsmath}
\usepackage{amsfonts}
\usepackage{color}
\usepackage{epsfig}

\def \half{{1 \over 2}}
\def \da{^{\dagger}}
\def \etal{{\it et al.}~}

\def \Z2{$\mathbb{Z}_2$}
\def \A{\mathcal{A}}
\def \nuvec{\boldsymbol{\nu}}
\def \hvec{\boldsymbol{h}}
\def \xvec{\boldsymbol{x}}
\def \kvec{\boldsymbol{k}}
\def \pvec{\boldsymbol{p}}
\def \qvec{\boldsymbol{q}}

\def \Qvec{\boldsymbol{Q}}

\begin{document}

% -------- Title -----------
\title{The strong side of weak topological insulators }

\author{Zohar Ringel}
\altaffiliation{Contributed equally to this work.}
\affiliation{Department of Condensed Matter Physics, Weizmann Institute of Science, Rehovot 76100, Israel}

\author{Yaacov E.~Kraus}
\altaffiliation{Contributed equally to this work.}
\affiliation{Department of Condensed Matter Physics, Weizmann Institute of Science, Rehovot 76100, Israel}

\author{Ady Stern}
\affiliation{Department of Condensed Matter Physics, Weizmann Institute of Science, Rehovot 76100, Israel}

% -------- Abstract -----------
\begin{abstract}
Three-dimensional topological insulators are classified into ``strong'' (STI) and ``weak'' (WTI) according to the nature of their surface states. While the surface states of the STI are topologically protected from localization, this does not hold for the WTI.
In this work we show that the surface states of the WTI are actually protected from any random perturbation that does not break time-reversal symmetry, and does not close the bulk energy gap. Consequently, the conductivity of metallic surfaces in the clean system remains finite even in the presence of strong disorder of this type. In the weak disorder limit the surfaces are found to be perfect metals, and strong surface disorder only acts to push the metallic surfaces inwards. We find that the WTI differs from the STI primarily in its anisotropy, and that the anisotropy is not a sign of its weakness but rather of its richness.
\end{abstract}

\maketitle

\section{Introduction}
\label{Sec:Intro}

Topological insulators (TIs) have recently become a very active subject in condensed matter physics. The classification of states of matter according to topological indices opens new horizons both theoretically and experimentally, and may hopefully lead to applications \cite{RMP_TI,RMP_TI2}.

For more than two decades, topological classification of phases was manifested primarily in the realm of the quantum Hall effect. The experimental observations of TI in two dimensions \cite{Konig} (2D) and three dimensions \cite{Hsieh} (3D) expanded this notion also to systems that are time-reversal (TR) symmetric, and have sparked a ``race for golf'' for new topological phases, and for their unique properties.

In 2D, TR symmetric band insulators are classified into ``trivial'' and ``topological'' by a \Z2 index \cite{KaneMele_Z2}. At the one-dimensional (1D) interfaces between a topological insulator and the vacuum (or any other trivial insulator), the energy gap must close, implying the appearance of counter-propagating chiral gapless modes. As long as TR symmetry is preserved, these modes are protected from back-scattering and gapping. In contrast, when a bi-layer system is formed of two such TIs, coupling of the edge modes in the two layers may gap them without violating the TR symmetry.

In 3D, TI's are classified by four \Z2-indices $(\nu_0, \nuvec)$ \cite{FuKaneMele, MooreBalents, Roy, Ashvin}. A non-trivial $\nu_0$ implies that on each 2D surface of the sample, the bulk gap is closed by surface states, the spectrum of which consists of an odd number of Dirac cones. As long as TR symmetry is preserved and the bulk gap remains open, at least one Dirac cone will survive the addition of any perturbation. Moreover, the wavefunctions of this Dirac cone can not be localized by disorder, and the surface of the 3D TI is apparently a perfect metal in the absence of electron-electron interaction \cite{Bardason_STI, Nomura_STI, Mirlin_graphene, Mudry}. Because of the robustness of its surface states, this phase was called ``strong TI'' (STI) \cite{FuKaneMele}.

On the other hand, if $\nu_0 = 0$ but $\nuvec \neq 0$, the system is in a phase known as a ``weak TI'' (WTI). This phase is adiabatically connected to stacked layers of 2D TI's \cite{FuKaneMele}. Suppose we have a cubic sample. The two surfaces which are aligned with the top and bottom layers will in general be gapped. But, the four perpendicular surfaces have gapless states, at least in a clean system. In the limit of completely decoupled layers, these surface states are actually the edge states of the stacked 2D TI. Translation-invariant coupling between the layers gaps out most of these surface states. However, Kramer's theorem ensures two Dirac cones to remain, both centered at momenta that are TR invariant. In the following, we refer to this type of surfaces, unless otherwise stated.

The chief reason why the WTI is considered weak is that its surface modes may be gapped without breaking TR symmetry or closing the bulk gap. In the stacked-layers picture, a mass term that gaps the edge modes arises if one couples the layers in pairs. The only symmetry violated by this term is the lattice-translation symmetry. Therefore, it appears that this symmetry is essential for the topological protection of the WTI surfaces. Since disorder breaks translational symmetry, one may be led to assume that the WTI surfaces are no longer protected and behave like conventional 2D metals with strong spin-orbit couplings. Such metals are known to undergo an Anderson transition from metals to insulators as a function of disorder strength \cite{Mirlin_review}.

In this paper, we show that the contrary is true. We consider the effect of disorder on the weak TI, and show that it is actually not weak at all. In Sec.~\ref{Sec:Finite}, we show that the conductivity of the non-trivial surfaces of the WTI remains higher than $e^2/h$ in the presence of disorder of arbitrary strength, as long as the bulk gap and TR symmetry are maintained.

Section \ref{Sec:Perturb} includes perturbative analysis. In the limit of weak disorder, we evaluate the weak localization correction, and find it to be anti-localizing. In the opposite limit, we consider strong disorder that is limited to several atomic layers at the surface of the insulator. We find that such disorder makes the surface insulating, but creates a perfect metallic sheet just beneath the disordered surface. We also discuss the conductivity in the intermediate disorder limit, and raise the possibility that a phase with a universal finite conductivity appears.

In light of these results, we discuss in Sec.~\ref{Sec:Anisotropy} the unique surface anisotropy of the WTI, which implies that the robustness of the conductivity of a surface strongly depends on its orientation. Based on this anisotropy we raise the possibility of surface engineering.

\section{Finite conductivity}
\label{Sec:Finite}

Assume one stacks an even number of 2D TI's, and couples them in pairs. Each such pair is topologically trivial, and generically has an insulating edge. Thus, in the 3D limit of an infinite even number of layers the surfaces are generically insulating. On the other hand, if the number of layers is odd, there is no way to gap all the edge modes without breaking TR symmetry, and the surface must be conducting. This sensitivity to the parity of the layer number was then argued to imply the fragility of the WTI~\cite{FuKane_Inversion}.

While the argument for non-triviality in the odd case relies on topology, the argument for gapping of the surface modes in the even case relies on a well-tailored perturbation that couples the layers in pairs. Random disorder does not induce such a coherent perturbation. Rather, when disorder is present and the number of layers is even, the surfaces \emph{may} be trivial, yet do not have to be. On the other hand, for an odd number of layers, the surfaces \emph{must} conduct. This suggests that when the coupling between layers is disordered, the odd behavior is in fact the generic one, thus the surfaces will conduct for any large number of layers.

\begin{figure}[tbh]
\begin{center}
\includegraphics[width=\columnwidth]{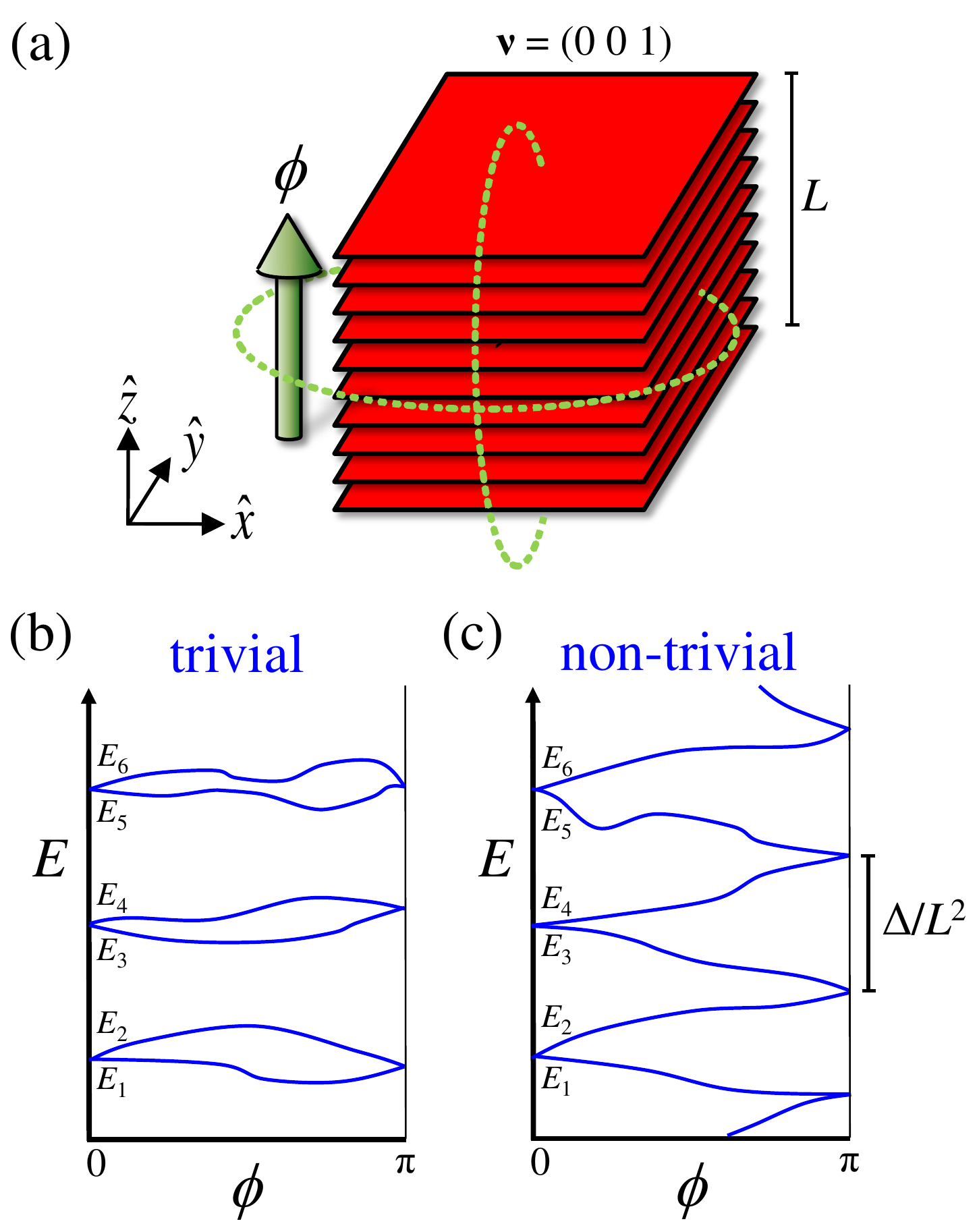}
\caption{ \label{Fig:switching} %
Topologically trivial and non-trivial pair switching. (a) The WTI is adiabatically connected to stacked layers of 2D TI. We consider a WTI of dimension $L^3$ with $\nuvec = (0\,0\,1)$ and $\hat{z}$ as the stacking direction. The boundary conditions (green dotted lines) are periodic in the $\hat{z}$ direction and are twisted by an Aharonov-Bohm flux in the $\hat{x}$ direction (green thick arrow). The remaining $xz$ surfaces are metallic. (b), (c) Typical patterns of energies of surface state as a function of the Aharonov-Bohm flux, $\phi$, for (b) trivial and (c) non-trivial surfaces. Kramer's theorem assures that at the time-reversal invariant fluxes $\phi = 0,\pi$, states come in degenerate pairs. On a trivial surface, the pairs remain the same between these two values, while on a non-trivial surface the pair switch partners. The mean level spacing of the surfaces states is $\Delta/L^2$, where $\Delta$ is the bulk gap. We show that non-trivial pair switching implies the existence of at least $O(L)$ extended states. }
\end{center}
\end{figure}

This heuristic argument will now be put on firm theoretical ground. Consider a WTI of dimension $L^3$ which is adiabatically connected to an odd number of 2D layers stacked along the $\hat{z}$ direction and $L \gg 1$. Note that we take the lattice spacing to be $1$ . We take the periodic boundary conditions to be periodic in the $\hat{z}$ and $\hat{x}$ directions, and open in the $\hat{y}$ direction, as illustrated in Fig.~\ref{Fig:switching}(a). Under these boundary conditions, the surface states reside on the interior and exterior surfaces of a thickened torus. We allow for any disorder which is TR symmetric, does not close the bulk gap, and has a correlation length much smaller than $L$. Under these conditions, the surfaces have no special regions or lines to which the electrons wave functions could be restricted.

Consider an Aharonov-Bohm flux that implements a phase twist $\phi$ in the periodic boundary conditions along the $\hat{x}$ direction, as illustrated in Fig.~\ref{Fig:switching}(a). Let us study how the spectrum of the edge modes depend on $\phi$. For $\phi=0,\pi$, the Hamiltonian is TR symmetric, and Kramer's theorem guarantees that all the energies are doubly degenerate. Apart from these degeneracies, the spectrum has no accidental degeneracies, as implied by the non-crossing theorem \cite{WIGNER}. This ensures a well-defined labeling of energies as a function of $\phi$, $E_i(\phi)$, where $i=1,2,\ldots$ and $E_{i+1} \ge E_i$.

The difference between topologically trivial and non-trivial surfaces is manifested in the relation between the pairs of degenerate states at $\phi=0$ and at $\pi$ \cite{FuKaneMele,EssinMoore,PAPER1}, which is illustrated in Fig.~\ref{Fig:switching}(b) and \ref{Fig:switching}(c). If the pairs at $\phi=0$ are the same as those at $\phi=\pi$, the surface is topologically trivial. This is the case for a trivial insulator and for a WTI with an even number of 2D layers. In contrast, if the pairs switch partners between $\phi=0$ and $\pi$, the surface is non-trivial. Such pair switching takes place on the surfaces of an STI and of a WTI with an odd number of 2D layers. The zigzag shape of the spectrum in the non-trivial surfaces cannot be terminated without approaching the bulk states. Hence, in this case
\begin{align}
\label{Eq:flow}
\sum_i |E_i(\pi)-E_i(0)| \geq \Delta,
\end{align}
where the summation is over all surface states and $\Delta$ is the bulk gap. Note that for any finite $L$, the number of surface states is proportional to $L^2$, and the mean level spacing is $\Delta/L^2$.

It is impossible to satisfy inequality (\ref{Eq:flow}) if all the surface states are exponentially localized. The current of a localized state is exponentially small with the system size. The current carried by an electron in the $i^\textrm{th}$ eigenstate is given by $I_i(\phi) = (e/h)\partial_\phi E_i$ \cite{BLOCH}. Therefore, $\partial_\phi E_i \sim e^{-L}$, and consequently, $|E_i(\pi)-E_i(0)| \sim e^{-L}$. In that case inequality (\ref{Eq:flow}) cannot be satisfied.

Furthermore, $I_i = e\langle v\rangle_i / L$, where $\langle v\rangle_i$ is the expectation value of the velocity. Since, the velocity is bounded by intrinsic variables and cannot increase with the system size, $I_i$ approaches zero at least as $1/L$. Note that there is a value $\phi_0$ such that  $|E_i(\pi)-E_i(0)| \leq (\pi h/e) |I_i(\phi_0)|$. Therefore, for the inequality (\ref{Eq:flow}) to be satisfied, there must be at least $O(L)$ delocalized states. Furthermore, as long as the system is homogeneous, which is the case for random disorder, these states are distributed all over the surface.

Imagine now cutting the system into two subsystems, one with even and one with odd number of layers. Since the cut is a surface effect and the system is homogeneous, it will not localize a state that has been delocalized before. Thus, in the presence of random disorder, delocalized states will exist also in the subsystem with the even number of layers. We can therefore conclude that in a system with an even number of layers there are delocalized states in the presence of random disorder, despite the absence of topological protection.

The homogeneity of the disordered system leads to a further consequence. Suppose we have cubic slabs of dimension $l^3$. According to what we have seen before, on the surface of the small cubes there are at least $O(l)$ delocalized states on the scale of $l$. Now we glue the cubes to one another and obtain a larger cube of dimension $L^3$. Since the gluing process does not localize states, on the surface of the large cube there are at least $O(L^2/l^2)$ delocalized states on the scale of $l$. This scaling is consistent both with delocalization of all states, and with a scenario of localized states with a broad distribution of localization lengths.

Finally, we notice that for the current $I_i$ to decay as $1/L$, the electronic motion must be ballistic. If, however, the motion is diffusive, then the current decays as $1/L^2$. In ballistic motion, inequality (\ref{Eq:flow}) required $O(L)$ delocalized states. In contrast,
a diffusive motion requires $O(L^2)$ such states. Since ballistic motion is unlikely in the presence of disorder, the bound of $O(L)$ states is probably too restrictive.

Having showed the existence of delocalized states, we can turn to estimate a lower bound for the conductivity. We use the Thouless formula, which relates the electrical conductivity to the sensitivity of energies to phase twists \cite{Thouless,GangOf4,Debney,Kohn},
\begin{align} \label{Eq:sigma_flux}:
\sigma_{xx} \approx \frac{e^2}{h} \langle \frac{\Delta E}{\Delta \phi} \rangle
\frac{dN}{dE},
\end{align}
where $\langle \Delta E / \Delta \phi \rangle$ denotes geometric mean of the energy difference $E_i(\pi)-E_i(0)$ averaged over eigenstates and $dN/dE$ denotes the density of states, both at Fermi energy. This relation has been shown to be only qualitatively correct \cite{LeeReview}. For example, in 1D systems the conductivity scales like $[\langle \Delta E / \Delta \phi \rangle (dN/dE)]^2$ \cite{AndersonLee}, and constants of order unity may appear \cite{Debney,Montam}. Moreover, discrepancies of $O(1)$ may appear if the relation is expressed with $\partial^2 E / (\partial \phi)^2$, rather than with $\Delta E / \Delta \phi$ \cite{Debney}. Nevertheless, when $\langle \Delta E / \Delta \phi \rangle (dN/dE)$ is of the order of unity, the conductivity is expected to be of the order of $e^2/h$.

In the non-trivial pair switching, see Fig.~\ref{Fig:switching}(c), the zigzag shape of the spectrum relates $\langle \Delta E / \Delta \phi \rangle$ to the energy levels spacing $E_{i+1}-E_i$. And since the level spacing is the inverse density of states, it leads to $\langle \Delta E / \Delta \phi \rangle (dN/dE) \ge 1$. Consequently,
\begin{align} \label{Eq:sigma_xx}
\sigma_{xx} \ge \frac{e^2}{h}.
\end{align}
We have therefore arrived at our key result: a non-trivial surface of a WTI will remain conducting even in the presence of random disorder.

Preliminary numerical work, reported in Appendix \ref{App:Numerics}, indeed shows that as the number of stacked layers increases, the even-odd difference diminishes, and both tend to lack of localization.

\section{Perturbative analysis}
\label{Sec:Perturb}

The topological argument allowed us only to bound the conductivity. More quantitative predictions can be given in the limits of weak disorder and strong surface disorder, where perturbative approaches can be utilized. Disorder is defined to be weak when $E_F\tau\gg 1$, where $E_F$ is the Fermi energy and $\tau$ is the mean free time. In this limit we evaluate the lowest-order quantum correction to the conductivity \cite{Review_localization}.

The low energy effective Hamiltonian describing the surface of a WTI in the clean limit consists of two decoupled Dirac cones
\begin{align} \label{Eq:Hk_WTI}
H(k_x,k_y) = v_0 ( k_x I^* \otimes s_x + k_y I \otimes s_y ).
\end{align}
For every value of $k_x$ and $k_y$ this is a $4\times 4$ matrix, spanned by a direct product of two Pauli spinors: $\mathbf{\tau}$ that denotes the Dirac cone, and $\mathbf{s}$ that denotes the electron spin. Here $v_0$ is the velocity characterizing the Dirac cones, and $I^*$ may be either the unity matrix $I$ or the Pauli matrix $\tau_z$, depending on the particular WTI considered. The corresponding TR operator is $T_W = I \otimes i s_y K$, where $K$ denotes complex conjugation. Accordingly, $T_W^{\phantom{w}2} = -1$. Notably, since under TR each Dirac cones is mapped to itself, in general their chiralities are unrelated, as well as the energy of the Dirac points.

Disorder adds to the Hamiltonian a sum of the form $\sum_{m,n}V_{mn}({\bf r})(\tau_m \otimes s_n)$, where the indices $m,n$ take the values $0,x,y,z$ and $\tau_0 = s_0 = I$. For the WTI only six terms are TR symmetric: $I \otimes I, \tau_z \otimes I, \tau_x \otimes I$ and $\tau_y \otimes \mathbf{s}$. The first three describe potential disorder, and the last three describe random spin-orbit scattering (note that the clean Hamiltonian already includes spin-orbit). Among the six, only the term $\tau_y \otimes s_z$ gaps the spectrum.

In Appendix \ref{App:WAL} we evaluate the lowest order quantum correction to the conductivity from the low energy Hamiltonian (\ref{Eq:Hk_WTI}), in the presence of all mentioned types of disorder \cite{FirstOrder}. We find this correction of be anti-localizing,
\begin{align} \label{Eq:dlogsigma}
\frac{d\ln \tilde{\sigma}_{xx}}{d\ln L} = -T_W^{\phantom{W}2} \frac{1}{2 \pi \tilde{\sigma}_{xx}} f_v^{\phantom{v}2} \left( 1 - \frac{f_e}{2} \right) > 0,
\end{align}
where $\tilde{\sigma}_{xx} = \sigma_{xx} (h/e^2) = E_F \tau f_v$. Furthermore, $2/3 < f_v < 2$ is the vertex correction, and $-1 \leq f_e \leq 1$ is a correction of the Cooperon, both are determined by the details of the disorder. Equation (\ref{Eq:dlogsigma}) implies that the conductivity flows towards a perfect metal, and $\tilde\sigma_{xx}$ increases logarithmically with the system size.

The Hamiltonian (\ref{Eq:Hk_WTI}) appears also in two other 2D systems, and it is instructive to elucidate the similarities and differences between these systems and the WTI. The first system is that of spinless electrons in graphene. For that system $I^*=\tau_z$, and $\mathbf{s}$ denotes the sublattice index. Accordingly, its TR operator is $T_G = \tau_x \otimes I K$ and $T_G^{\phantom{g}2} = 1$. By plugging $T_G$ instead of $T_W$ into Eq.~(\ref{Eq:dlogsigma}), we observe in graphene weak localization, as expected \cite{AleinerEfetov,McCann}. As a matter of fact, since for the WTI $T_W^{\phantom{w}2}=-1$, for generic disorder the Hamiltonian belongs to the symplectic class, which is known to have weak anti-localization correction \cite{Mirlin_review}. In contrast, spinless graphene, for which $T_G^{\phantom{g}2}=1$, belongs to the orthogonal class, which shows weak localization. The general relation between the symmetry class and the sign of the weak-localization correction can be shown using the non-linear $\sigma$ model approach \cite{Mirlin_review}, but may be also understood more directly by means of interference of diffusive trajectories, as shown in Appendix \ref{App:Trajectories}.

The second system is that of a 2D insulator at the transition point between a trivial and a topological phase in the absence of inversion symmetry \cite{QSHETransition}. This system belongs to the same symmetry class as the WTI, but its spectrum does not exhibit pair switching as a function of flux. Nonetheless, since it is tuned to a phase transition, one expects the correlation length to diverge and therefore delocalized states must exist at low energies. In Refs.~\cite{Onoda,Hastings,Z2Network} it was established that a band of delocalized states appears around zero energy, while far from zero energy states are localized. This should be compared with the WTI, where, as we have argued, states remain delocalized for any sub-gap energy. This discrepancy suggests that while these models have similar low energy descriptions, they are nonetheless different
when the entire spectrum is taken into account.

Notably, following the posting of this manuscript on the arXiv, our prediction for delocalization within the low energy theory, Eq.~(\ref{Eq:Hk_WTI}), was validated numerically \cite{Mong}. The restriction of this numerical work to low energy does not address directly the role of pair switching. Indeed, one can find a unitary transformation that maps the low energy Hamiltonian used there, including the disorder terms, to the low energy part of the Hamiltonian used in Ref.~\cite{QSHETransition} to describe the transition mentioned in the previous paragraph, which exhibits no pair-switching.

% ---- Strong disorder ----
\vspace{5mm}

Having shown that in the limit of weak disorder we have a perfect metallic surface, we now turn to opposite the limit of extremely strong disorder. In this limit the strength of the disorder is much larger than all the other energy scales, including the bulk band width and band gap. If such disorder acts on the entire 3D system, it mixes the bulk bands and makes the entire sample a trivial insulator. However, an interesting case is disorder that is limited to several of the outmost layers. This may actually happen in realistic surfaces, which are usually made dirty by oxides and other dopants. Moreover as we show below, it also reveals the role of the bulk in protecting the surface states.

\begin{figure}[t]
\begin{center}
\includegraphics[width=\columnwidth]{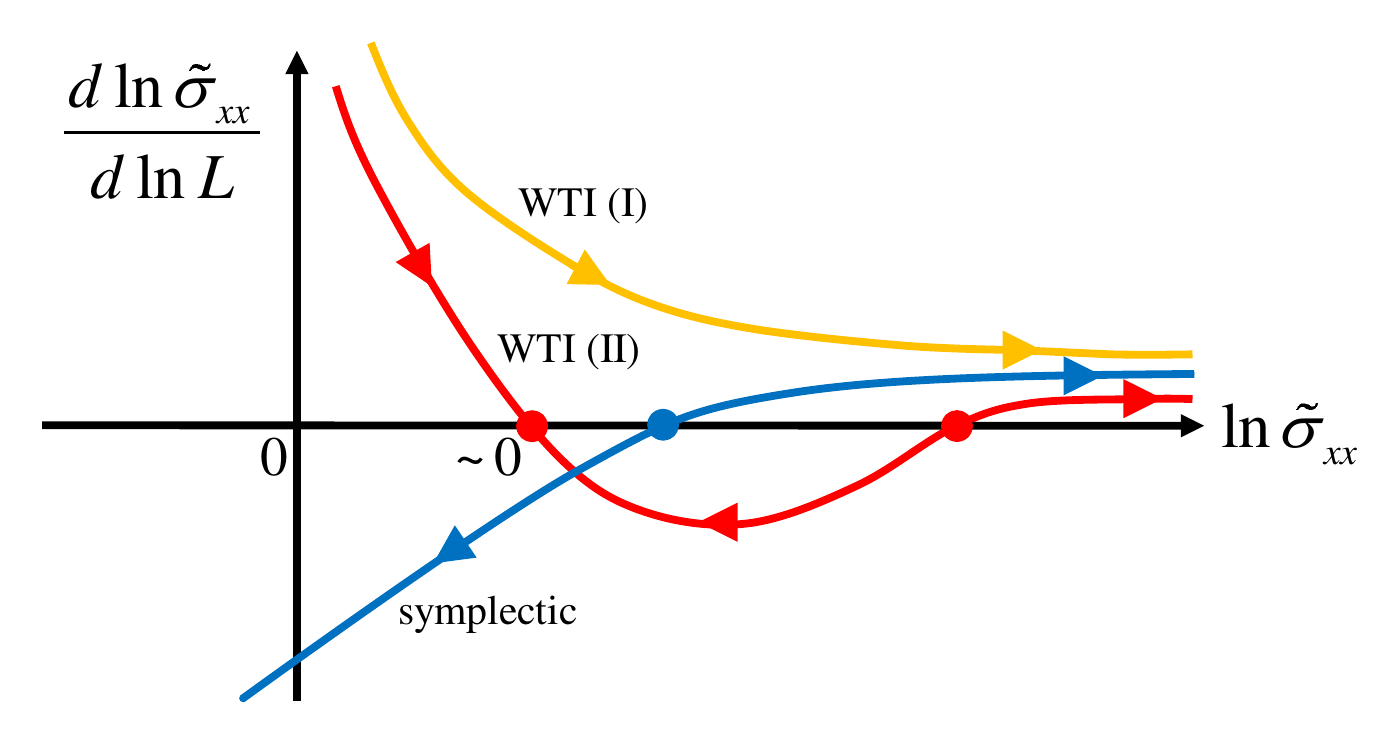}
\caption{ \label{Fig:phase_diagram} %
Renormalization group flow of the conductivity. The $\beta$-function of the dimensionless conductivity $\tilde{\sigma}_{xx}$ for a non-trivial surface of the WTI, compared with that of a 2D metal with strong spin-orbit coupling that belongs to the symplectic class (blue). According to Eq.~(\ref{Eq:dlogsigma}) in the limit of high conductivity the flow is toward a perfect metal. We showed that the conductivity of a WTI surface cannot drop below $e^2/h$. Consequently, two types of flows are possible: (I) always flowing towards a perfect
metal (yellow), and (II) flowing with a stable fixed point of finite conductivity (red).}
\end{center}
\end{figure}

Let us divide the Hamiltonian of the three dimensional system, $H_{\textrm{3D}}$, into the part that operates only within the clean bulk ($H_0$), the part that operates on the disordered surface layers ($H_\text{dis}$), and the part of hopping between the two ($V$). The Hamiltonian may now be written as
\begin{align} \label{Eq:H_matrix}
H_{\textrm{3D}} = \left( \begin{array}{cc} H_0 & V \\ V\da & H_\text{dis} \end{array} \right).
\end{align}
We begin with the case where all the eigenvalues of $H_\text{dis}$ are greater in absolute value than some value $W$, and $W\gg t$, where $t$ is the bulk band width. For this case, all the eigenstates of $H_\text{dis}$ are localized on the surface. These eigenstates may be considered as an high energy sector, which can be integrated out. To this end, we consider the Green's function projected onto the Hilbert space of the clean bulk  \cite{Assa}, using the projection operator $P_0$
\begin{eqnarray} \label{Eq:P0_G_P0}
& & P_0(E-H_{3\textrm{D}})^{-1}P_0 = \\
& & \qquad \left(E - H_0 - V (H_\text{dis}-E)^{-1} V\da + O(V^4)\right)^{-1}.  \nonumber
\end{eqnarray}
This Green's function defines an effective Hamiltonian for the clean bulk, which is
\begin{align} \label{Eq:Heff}
H_\text{eff}(E) = H_0 + V (H_\text{dis}-E)^{-1} V\da + O(V^4).
\end{align}
This effective Hamiltonian describes the degrees of freedom of a 3D WTI which is clean in the bulk (the first term), and is disordered at its surface (the second term). Note that this surface lies beneath the physical surface, where the disorder vanishes. The second term of Eq.~(\ref{Eq:Heff}) represents virtual hopping from the bulk to the strongly localized states at the physical surface and back. Since all the eigenvalues of $H_0$, the matrix elements of $V$ and the energy $E$ are of the order of $t$, this bulk-surface coupling is of the order $t^2/W \ll t$. The effective Hamiltonian then describes a {\it weakly disordered} WTI, the gapless states of which are located underneath the physical surface. Recalling the above result, we can see that the relocated surface states form a perfect metal.

The same holds when the spectrum of $H_\text{dis}$ becomes continuous. The states in the strongly disordered layers with energy greater than $W$ can still be integrated out, resulting in small $O(t^2/W)$ terms. The remaining low lying states are expected to be localized, since $H_\text{dis}$ alone is not protected from localization. Such states act as strong scatterers. However, their density is of $O(t/W)$, and is therefore small, yielding a long mean-free path. Hence, we are still in the limit of weak disorder, thus having a perfect metal.

\vspace{5mm}

Intermediate disorder is disorder with $E_F\tau\ll 1$ but $\Delta\tau \gg 1$. According to the topological argument, the conductivity has to be larger than $e^2/h$ even in this regime. Following the single parameter scaling approach, two possible flows of the renormalization group may arise, which can be presented in terms of the $\beta$-function, $\beta(\tilde{\sigma}_{xx}) = d\ln \tilde{\sigma}_{xx} / d\ln L$. In one flow, the conductivity always flows to infinity while increasing the system size, as presumably happens in the STI \cite{Bardason_STI, Nomura_STI, Mirlin_graphene, Mudry}. In the second flow, a stable fixed point appears at $\tilde{\sigma}_{xx} \approx 1$, and a critical point appears for some $\tilde{\sigma}_{xx} > 1$. The two flows are illustrated in Fig.~\ref{Fig:phase_diagram}. \cite{Flow_Mong}

\section{Surface anisotropy}
\label{Sec:Anisotropy}

In the previous sections we analyzed the conduction properties of surfaces of the WTI, and found that they are conducting even in the presence of disorder. This robustness brings the WTI closer to the STI in terms of their transport properties. Nevertheless, the WTI differs from the STI in the unique anisotropic behavior of its surfaces, which gives rise to the idea of surface engineering.

\begin{figure}[tb]
\begin{center}
\includegraphics[width=7cm]{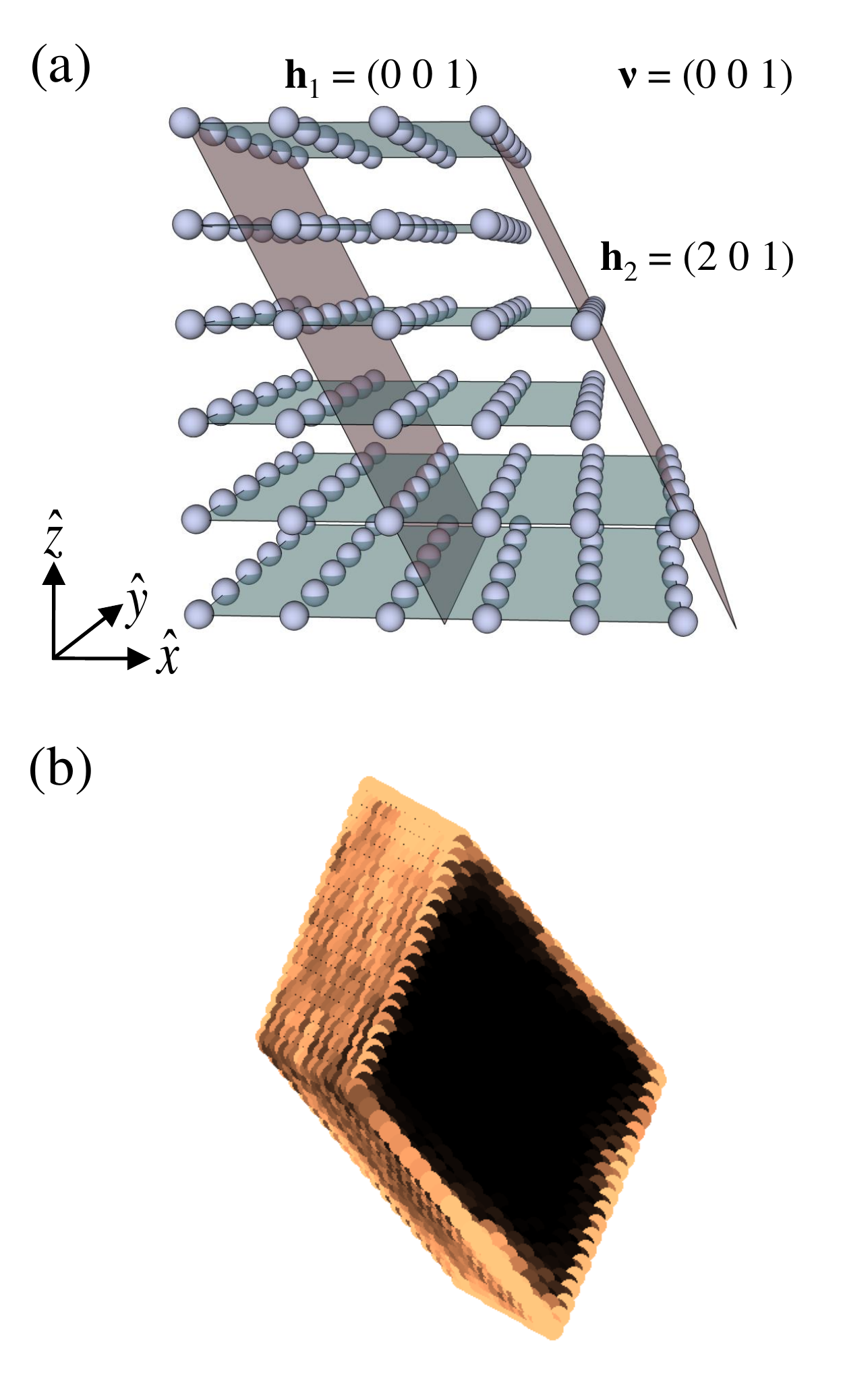}
\caption{ \label{Fig:cubes} %
Insulating and metallic surfaces. (a) The WTI has both trivial and non-trivial surfaces. A surface with Miller indices $\hvec$ is trivial if $(\hvec - \nuvec) \mod \, 2 = 0$, denoted by $\hvec \sim \nuvec$, since any such $\hvec$ can denote a stacking direction. The figure depicts two trivial surfaces $\hvec_1 = (0\,0\,1)$ and $\hvec_2 = (2\,0\,1)$ for a cubic crystal with assumed $\nuvec = (0\,0\,1)$. Both $\hvec_1$ and $\hvec_2$ are legitimate stacking directions. Alternatively, for a stacking along $\hvec_1$, the $\hvec_2$ surface is composed of steps of two layers. The coupling between the layers gaps theirs edge states. For $\hvec \nsim \nuvec$, the steps will be of odd number of layers and will therefore conduct. (b) An example of the surface anisotropy in the Fu, Kane and Mele model of the weak $\nuvec = (0 \, 0 \, 1)$ phase. Depicted is the local density of surface states integrated over an energy window $|E| < 0.1\Delta$, with disorder strength comparable to the bulk gap. The surfaces of the parallelepiped are spanned by the primitive vectors. The two faces with Miller indices equal to $\nuvec$ are gapped, while the other four, with orthogonal Miller indices, are metallic. By controlling the cleavage process, the conductance of each face of the WTI can be engineered. }
\end{center}
\end{figure}

While non-trivial surfaces of the WTI are indeed robustly conducting, not all possible surfaces of the WTI are topologically non-trivial. For example, we mention that in the stacked-layers picture the top and bottom surfaces are topologically trivial, and are generally gapped. For given weak indices $\nuvec$ and a plane with Miller indices $\hvec$, we define the relation $\hvec \sim \nuvec$ by
\begin{align} \label{Eq:h_mod_nu}
(h_i - \nu_i) \mod \, 2 = 0,
\end{align}
for $i = 1,2,3$. Any surface with Miller indices that satisfy this relation is topologically trivial, whereas a surface with $\hvec \nsim \nuvec$ is topologically non-trivial \cite{FuKaneMele}. The reason of this criterion is that the indices vector $\nuvec$ does not uniquely define a stacking direction, and any vector $\hvec \sim \nuvec$ can be a stacking direction, as illustrated in Fig.~\ref{Fig:cubes}(a). Namely, the stacked-layers picture is a theoretical construction rather than a physical description, and in practice, the WTI does not have to be layered.

An alternative explanation for criterion (\ref{Eq:h_mod_nu}) can be given from the picture of a fixed stacking direction and varying surfaces. Consider a WTI, the primitive lattice vectors of which are $\boldsymbol{a}_1, \boldsymbol{a}_2, \boldsymbol{a}_3$, making the lattice sites located at $\boldsymbol{r}_{\boldsymbol{n}} = \sum_{i=1}^3 n_i \boldsymbol{a}_i$. Consider also a surface with Miller indices $\hvec$, and for simplicity place the origin of the coordinate system at some lattice site of the surface. By definition, all the lattice sites on the surface satisfy the condition $\hvec \cdot \boldsymbol{n} = 0$. For simplicity, we take the example of $\nuvec = (0 \, 0 \, 1)$ and choose it to be the stacking direction. If $\hvec \sim \nuvec$, then $h_3$ is odd, while $h_1$ and $h_2$ are even. Accordingly, on the surface all the $n_3$ coordinates are even, and adjacent surface sites differ by an even increment of $n_3$. Therefore, the surface is composed of steps of an even number of layers, as illustrated in Fig.~\ref{Fig:cubes}(a), and the coupling between them will gap the edge states. For $\hvec \nsim \nuvec$, the surface is composed of steps of odd layers. Now, the coupling can not gap all the edge states, and the surface will conduct.

The high and non-trivial sensitivity of the surfaces to their orientation even in the presence of disorder is demonstrated in Fig.~\ref{Fig:cubes}(b). We considered a $20 \times 20 \times 20$ lattice of the Fu, Kane, and Mele model with $\lambda_{SO} = t$ and $\delta t = [-0.6,0,0.2,0]t$, which corresponds to $\nu_0 = 0$ and $\nuvec = (0 \, 0 \, 1)$ with a bulk gap of $\Delta = 0.8t$. In this model $\nuvec$ represents also the weak hopping direction. Uniformly distributed strong disorder of magnitude $t$ was also introduced. The figure depicts the local density of surface states integrated over an energy window $|E| < 0.1\Delta$. The parallelepiped is cut along the primitive vectors, and therefore has 2 trivial gapped faces and 4 topological metallic faces.

The criterion for a surface $\hvec \sim \nuvec$ implies that the spectrum on it will be gapped, but it does not provide information on the magnitude of the gap. In the above example, for $\hvec$ chosen to be in the weakest hopping direction, the energy gap on the surface is comparable to the bulk gap. Other trivial surfaces have energy gaps much smaller than this value. The influence of disorder on the gap and localization length of such surfaces may be dramatic. We note that for a surface that cannot be described by Miller indices, we expect metallic behavior, since the scaling argument which was used to ensure $\sigma_{xx} \geq e^2/h$ seems to hold.

By noticing that the topological and trivial surfaces are isotropically distributed, one can imagine creating a sample with each face engineered to be either gapped or metallic. A gapped surface along a stacking direction would remain insulating, while other surfaces will conduct. Provided rather good control on the cleaving process, various different electronic behaviors are expected on different surfaces, ranging all the way from perfect metals to insulators with varying gaps. In light of these results, we find that the anisotropic behavior of the WTI surfaces is not a sign of their weakness, but rather of their richness.

\section{Summary}

In this work, we showed that the name ``weak topological insulators'' does not do justice to the phase it describes, since the electrical conductivity of the non-trivial surfaces of such insulators is not suppressed by disorder. The WTI shows unique sensitivity of the electronic properties of its surfaces to their orientation, and that may provide an experimental tool for controlling these properties. We hope that this work will serve as a trigger for further study of these interesting topological phases.

\section*{Acknowledgements}
The authors thank Y Imry, FDM Haldane and IA Gruzberg for useful discussions. ZR thanks ISF grant 700822030182. YEK and AS thank the US-Israel Binational Science Foundation, the Minerva foundation and Microsoft's station Q for financial support.

%\section*{Author contributions}
%ZR and YEK contributed equally to this work.

\appendix
\section{Numerical analysis of disordered thin WTI}
\label{App:Numerics}

In an attempt to address numerically the effect of disorder on the conductivity of a gapless surface of the WTI, we considered the Fu, Kane and Mele model \cite{FuKaneMele} with $\lambda_{SO} = t$ and $\delta t = [-0.6,0,0.2,0]t$, which corresponds to a $\nu_0 = 0$ and $\nuvec = (0 \, 0 \, 1)$ with a bulk gap of $\Delta = 0.8t$. We also took the chemical potential to be at the Dirac points of the surface spectrum. The most general potential disorder that is symmetric to time-reversal was included by adding a time reversal symmetric random matrix which acts within unit cells. The entries of each matrix were sampled from a uniform distribution in some region $[-w/2,w/2]$, and the resulting matrix was then symmetrized with respect to time reversal. The disorder was added on three outmost layers with $w = 0.5t,0.5 e^{-2}t,0.5e^{-3}t$ corresponding to the first, second and third layer respectively. The samples sizes $L_x \times L_y \times L_z$ ranged in from $40 \times 10 \times 1$ up to $120 \times 10 \times 6$ unit cells, where $L_z$ can be thought as the number of the stacked 2D layers.

In order to obtain the conductance $g_{xx}$ we used Eq.~(2) of the main text. When applied to a quasi-1D sample this equation yields the conductance rather than the conductivity \cite{GangOf4}. The fluctuations in energy levels following the insertion of a $\pi$-twist were approximated by extrapolating the derivative of the energy levels with respect to the phase twist. The geometric averaging was taken over the different instances of disorder and over an energy window of $[-0.2t,0.2t]$. Although this second averaging is not included in the definition, we find that it did not have significant influence on the asymptotic behavior. We considered 30 instances of disorder for 1-3 layers, and 10 instance of disorder for 4-6 layers. The error bars are primarily due to fluctuations of the density of states which limit the accuracy of the estimated mean value.

\begin{figure}[tbh]
\begin{center}
\includegraphics[width=8cm,height=6cm,angle=0]{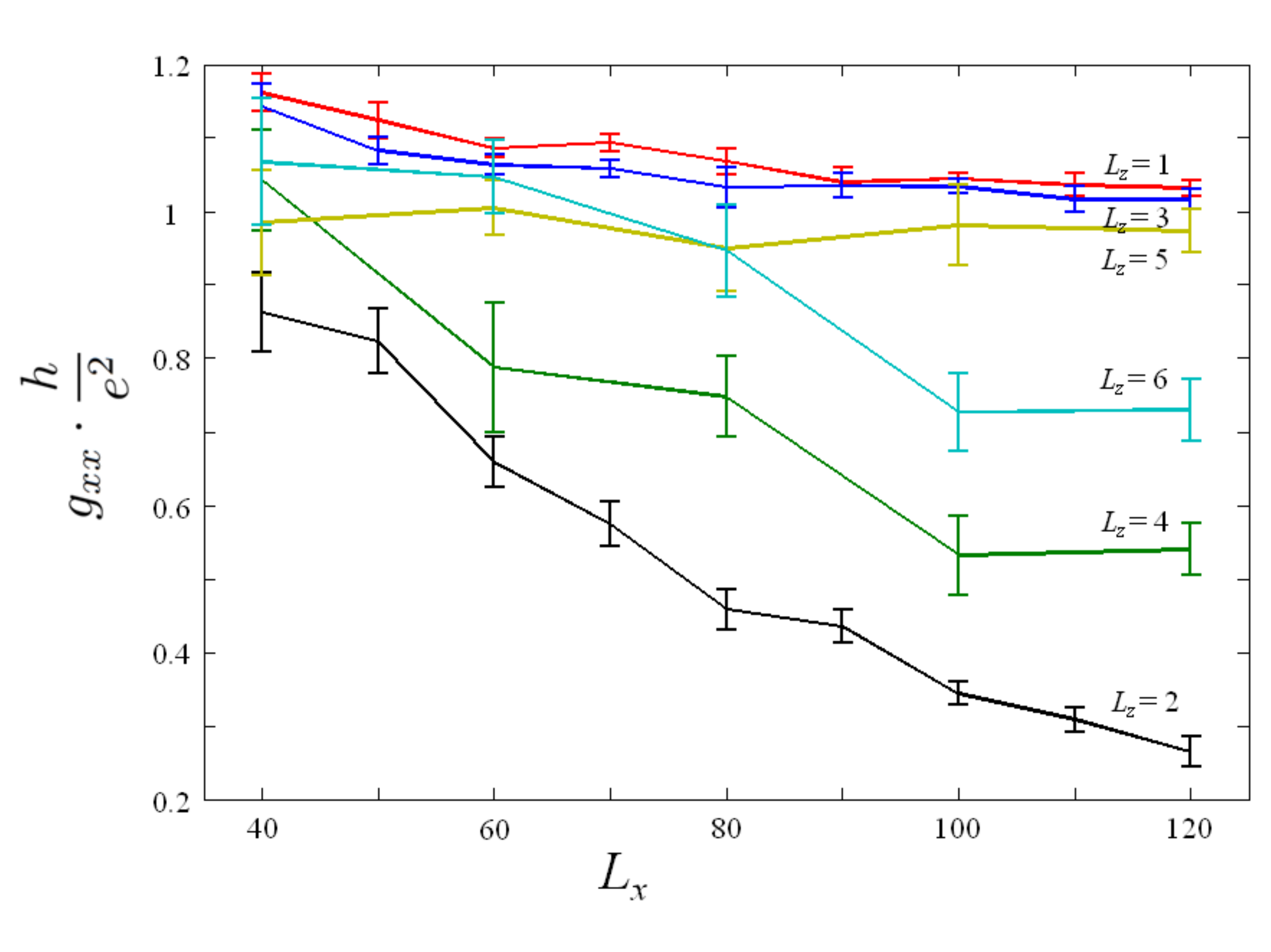}
\caption{ \label{Fig:GofLandN} %
The conductance $g_{xx}$ for a WTI with $\nuvec = (0 \, 0 \, 1)$ of the Fu, Kane and Mele model as estimated from the flux sensitivity of surface state energies multiplied by the density of states. Each line corresponds to a given number of layers ($L_z$) and shows $g_{xx}$ as a function of $L_x$, where $L_y$ is fixed to 10. Periodic boundary conditions were imposed on $\hat{z}$ and $\hat{x}$. Samples with an odd number of layers have a topologically protected minimal conductance of $e^2/h$. Samples with an even number of layers are (strictly speaking) topologically trivial and show a localization behavior. However their conductance converges to that of the odd layers as the number of channels is increased.}
\end{center}
\end{figure}

The dependence of $g_{xx}$ on the dimensions of the surface is depicted in Fig.~\ref{Fig:GofLandN}. For odd $L_z$ the conductance tends to values close to the $e^2/h$, and shows no sign of localization. For comparison, the localization length of a sample with disorder of a similar strength that does not satisfy time reversal symmetry is around 40 unit cells. For even $L_z$ a finite localization length is apparent, which however increases with $L_z$. It therefore appears that for large $L_z$ the even curves will converge to the odd curves, meaning a lack of localization for $L_z \gg 1$.

We note that a similar behavior was obtained in Ref.~\cite{Takane} for symplectic multichannel 1D wires. This model is close to ours, but with one important difference. In multichannel 1D wires all the channels are coupled, while in the 2D surface of WTI only nearby channels are coupled.

\section{The first order quantum corrections to the conductivity}
\label{App:WAL}

In this appendix we derive the lowest order quantum corrections to the electrical conductivity of WTI and spinless graphene. While the former is our main interest, we find it instructive to compare it to the latter. Our starting point is a low energy effective Hamiltonian for both systems. The Hamiltonian is composed of two decoupled Dirac cones
\begin{align}
H_0 &= -i v_0 (\partial_x I^* \otimes s_x + \partial_y I \otimes s_y),
\end{align}
where $s_i$ are Pauli matrices associated with the spin (sublattice) index of WTI (graphene), and the matrix $I^{*}$ is a Pauli matrix associated with the valley index (cf.~Eq.~(4) in the main text). For WTI, $I^*$ denotes either $I_{2x2}$ or $\tau_z$, while for graphene $I^* = \tau_z$. The corresponding retarded and advanced Green's functions are given by
\begin{align}
G_0^{R/A}(\kvec,E) &= \frac{E + v_0 k_x I^* \otimes s_x + v_0 k_y I \otimes s_y}{E_\pm^{\phantom{\pm}2} - (v_0 k)^2},
\end{align}
where $E_\pm = \lim_{\eta \rightarrow 0^+} E \pm i \eta$. Time reversal invariant potential disorder is introduced via the matrix $V(\xvec)$
\begin{align}
H &= H_0 + V(\xvec), \\
V(\xvec) &= \sum_l v_l(\xvec) A_l,
\end{align}
where $A_l$ are $4 \times 4$ time-reversal symmetric Hermitian matrices of the form $\tau_i \otimes s_j$ for $i,j = 0,x,y,z$. The $v_l(\xvec)$ are uncorrelated random functions
\begin{align} \label{Eq:vlvl}
\langle v_l(\xvec) v_{l'}(\xvec')\rangle = w_l \delta_{ll'} \delta(\xvec - \xvec').
\end{align}

As mentioned in the main part of the paper, the time-reversal operator $T$ is different for spinless graphene and WTI. For spinless graphene, the time-reversal operator switches between the two Dirac points, but does not affect the sublattice. Therefore, $T_g = \tau_x \otimes I K$, where $K$ denotes complex conjugation. On the other hand, for WTI it flips the spins but does not affect the valleys, since the Dirac points are at time-reversal-invariant momenta. Therefore, $T_W = I \otimes s_y K$. Consequently, $T_W^{\phantom{W}2} = -1$ while $T_g^{\phantom{g}2} = 1$. As argued in the main work using the particle diffusion picture, the signs of the quantum interference correction to the conductivity is expected to be given by $-T^2$. Another consequence of the difference in $T$ is that the $A_l$ matrices which commute with $T_g$ are all the combinations of $(I,\tau_x,\tau_y) \otimes (I,s_x,s_z)$ and $\tau_z \otimes s_y$, while the matrices which commute with $T_W$ are $I \otimes I, \tau_x \otimes I, \tau_z \otimes I, \tau_y \otimes s_x,\tau_y \otimes s_y, \tau_y \otimes s_z$.

Due to extra symmetries of $H_0$, there are additional anti-unitary operators which commute with $H_0$. For example, for $H_0$ in which $I^* = I$, all the $\tau_i T_W$ matrices are such operators. If one chooses disorder that commutes with $\tau_i T_W$, rather than with $T_W$, then the sign of the quantum correction will be $-(\tau_i T_W)^2$.

Our goal is to find the changes in the disorder-averaged conductance as a function of the linear size of the system. The zero-temperature mean longitudinal conductance in the $x$ direction is given by \cite{McCann}
\begin{align}
\sigma_{xx} &= \frac{e^2}{2\pi \hbar} \left \langle \int \frac{d^2 p}{(2\pi)^2} \textrm{Tr} [J_x G^R(\pvec,E_F) J_x G^A(\pvec,E_F)] \right \rangle, \nonumber
\end{align}
where $\langle.. \rangle$ denotes averaging over disorder, and $J_x = v_0 I^* \otimes s_x$ is the current operator.
The diagrammatic way to find this mean value combines the disorder averaged Green's function, the vertex correction, the Cooperon and the dressed Hikami box.

Our derivation follows McCann \etal in Ref.~\cite{McCann} for spinless graphene, but with three substantial differences. First, we address here both the WTI and graphene simultaneously in a way that emphasizes the differences between them, both in the Hamiltonian and in the resulting correction. Second, the only assumption we make on the disorder is that it is symmetric with respect to time reversal. We do not assume a dominance of one type of scattering over another.  Consequently, the numerical prefactor of the $\beta$ function depends on the details of the disorder, and these details may affect it by a factor of up to $1/3$. Last, since the spectrum of WTI far from the Dirac point is not universal, we adopt a different regularization approach for diverging integrals. Instead of introducing a triangular wrapping, we limit the minimal length scale of the scatterers. For alternative approaches for dealing with this issue see Refs.~\cite{AleinerEfetov,Mirlin_warp}.

\vspace{5mm}
% ---- Self-energy ----

We begin with calculating the self-energy within the self consistent Born approximation, given by
\begin{align}
\label{Eq:Sigma1-I}
\Sigma^R_1(\qvec,E)  &= \sum_l w_l \int \frac{d^2 p}{(2\pi)^2} A_l G^R A_l.
\end{align}
Since $A_l^{\phantom{l}2} = I \otimes I$, and the angular integration over $\pvec$ leaves only the diagonal term in $G^R$,
\begin{align} \label{Eq:Sigma1-II}
[\Sigma^R_1]_{ij}(\qvec,E)  &= \delta_{ij} \left[ i \Gamma + \frac{2\Gamma}{\pi} \ln(\frac{v_0 \Lambda}{E}) \right].
\end{align}
where $i,j = 1..4$. In the limit of weak disorder
\begin{align} \label{Eq:Sigma1-II}
\Gamma &= \frac{(\sum_l w_l) E}{4 v_0^{\phantom{0}2}}.
\end{align}
The level width $\Gamma$ is related to the mean free time $\tau$ through $\tau = 1/2\Gamma$. In order to obtain Eq.~(\ref{Eq:Sigma1-II}) we have introduced an ultra violet cutoff $\Lambda$, which physically corresponds to the characteristic inverse size of the impurities, and we assumed that $v_0 \Lambda$ is much smaller than the bulk gap (in WTI) or the bandwidth (in graphene). In the following we ignore the real part of the self-energy, since it only corresponds to a shift in the energy. The disorder averaged Green's function is now given by
\begin{align}
G^{R/A}(k,E) \approx \frac{E \pm i \Gamma + v_0 k_x I^* \otimes s_x + v_0 k_y I \otimes s_y}{(E \pm i \Gamma)^2 - (v_0 k)^2}. \nonumber
\end{align}

\begin{figure}[htb]
\begin{center}
\vspace{0cm}
\includegraphics[width=9cm,angle=0]{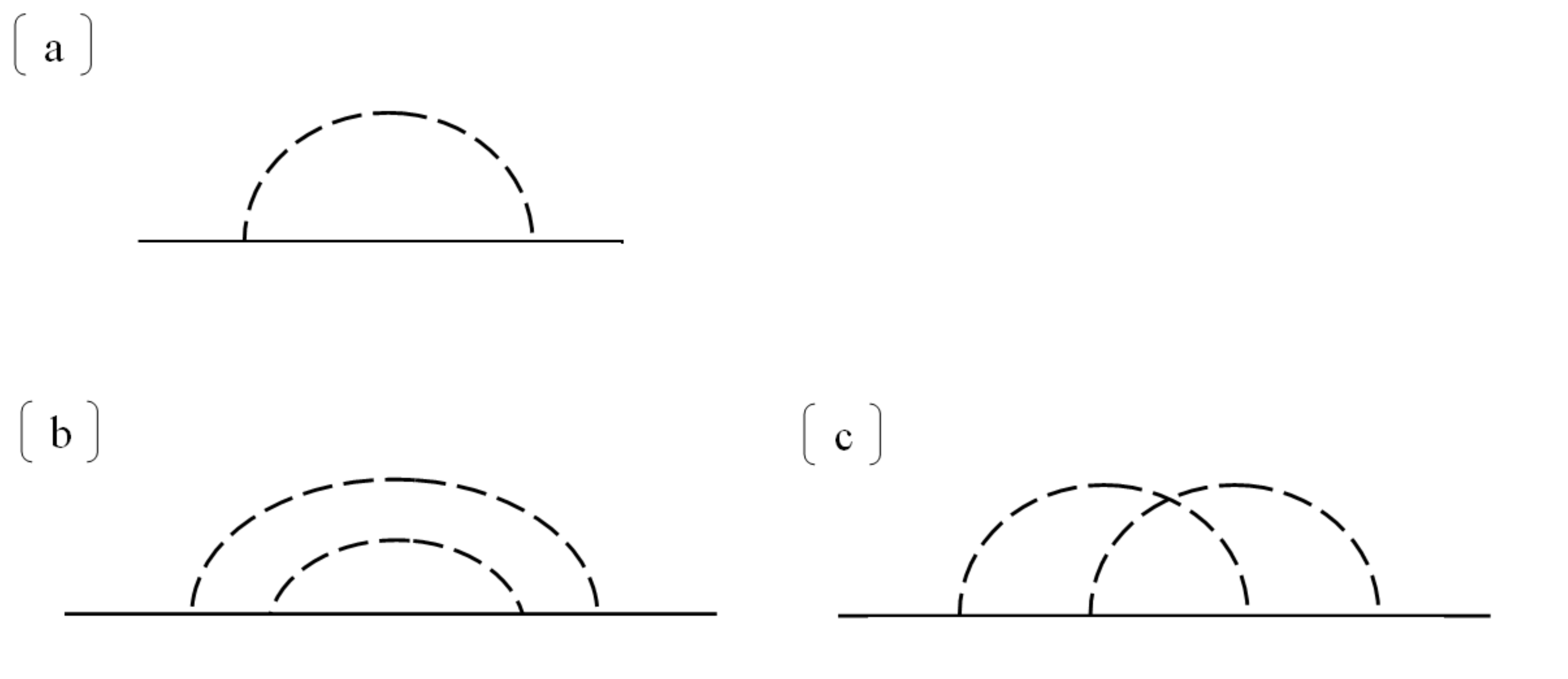}
\vspace{0cm} \caption{ \label{Fig:SelfEnergy} %
The leading self-energy diagrams in the Born approximation: (a) first order, and (b), (c) second order.}
\end{center}
\end{figure}

The self-consistent Born approximation includes diagrams where disorder lines do not intersect, such as depicted in Figs.~\ref{Fig:SelfEnergy}(a) and \ref{Fig:SelfEnergy}(b). It leaves out diagrams where disorder lines intersect, such as Fig.~\ref{Fig:SelfEnergy}(c). We find the self-consistent Born approximation to be valid when
\begin{align} \label{Eq:Limit}
\frac{\Gamma}{E} \ll 1 \quad \textrm{and} \quad \alpha \equiv \frac{\Gamma}{E} \ln\frac{v_0 \Lambda}{E} \ll 1.
\end{align}
Note that due to the logarithmic factor, even when $v_0\Lambda$ becomes much larger than $E$ there is still a wide parameter range in which the conditions are satisfied. Moreover, within this range the omission of the real part of the self energy is consistent.

In the limit of weak disorder diagram of Fig.~\ref{Fig:SelfEnergy}(b) is included in our approximation, but its contribution is smaller by a factor of $\alpha$ relative to that of Fig.~\ref{Fig:SelfEnergy}(a). Its contribution is
\begin{align} \label{Eq:Sigma_2a}
\Sigma^R_{2a} & (\qvec,E) = \sum_{l,l'} w_{l}w_{l'} \int \frac{d^2 p_1 d^2 p_2}{(2 \pi)^4} \times \\
& A_l G_0^R(\pvec_1,E) A_{l'} G_0^R(\pvec_1 - \pvec_2,E) A_l G_0^R(\pvec_1,E) A_{l'}. \nonumber
\end{align}
Due to the nested structure of the diagram, the integration over the two loops can be carried separately. The contribution of the diagram is therefore
\begin{align} \label{Eq:Sigma_2a_res}
\Sigma^R_{2a} \sim (\Gamma/E_F)^2 \ln^2(v_0 \Lambda / E_F) \sim \alpha^2.
\end{align}
Indeed this contribution is negligible for $\alpha \ll 1$.  The crossed diagram, which is depicted in Fig.~\ref{Fig:SelfEnergy}(c), can also be shown to be of $O(\alpha^2)$.

\vspace{5mm}
% ---- vertex correction ----

\begin{figure}[t]
\begin{center}
\vspace{0cm}
\includegraphics[width=9cm,angle=0]{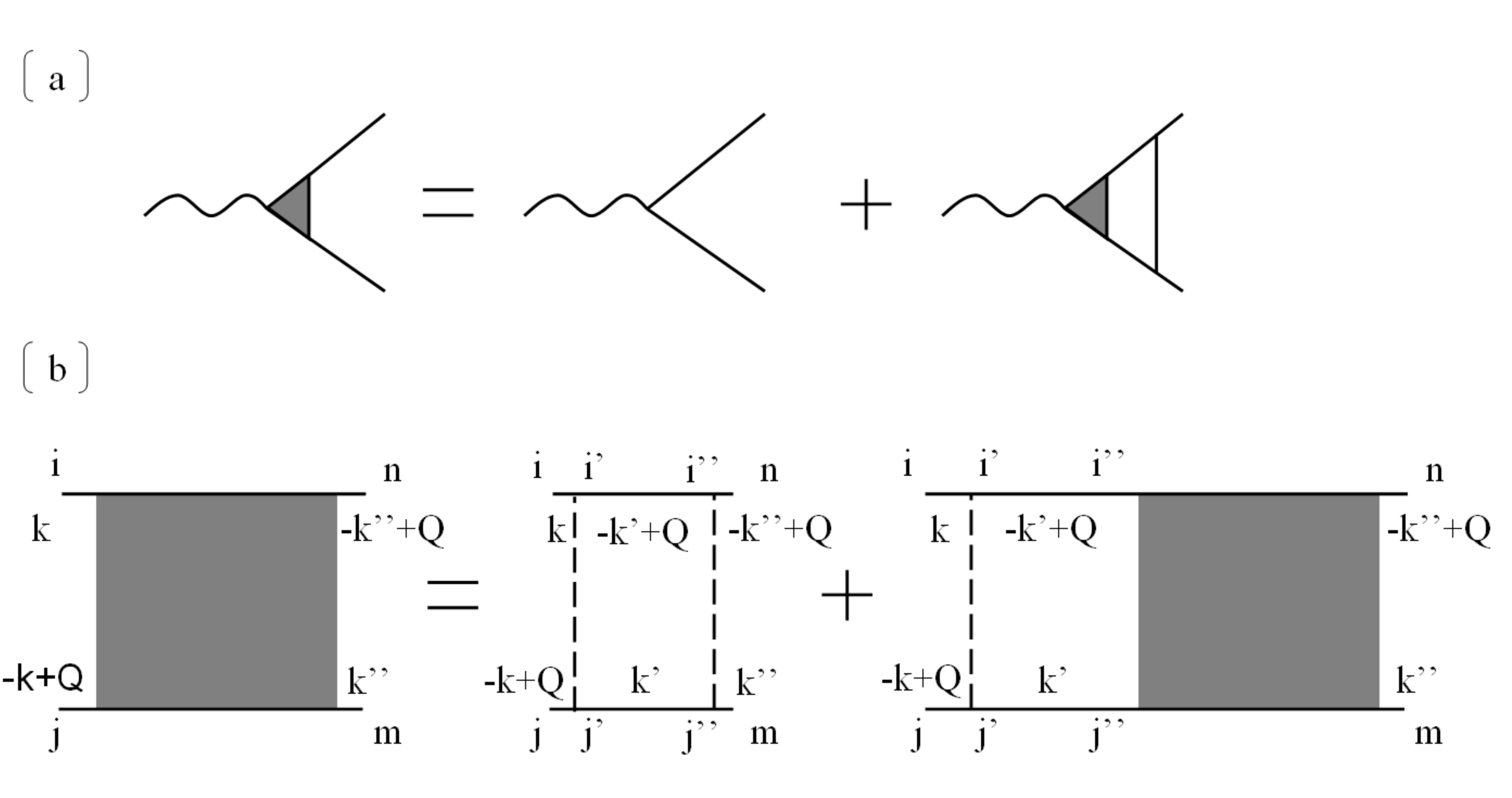}
\vspace{0cm} \caption{ Diagrammatic representation of the self-consistent equation for (a) the dressed vertex and (b)
the Cooperon. \label{Fig:SelfConst} %
}
\end{center}
\end{figure}

The self-consistent equation of the vertex correction is schematically illustrated in Fig.~\ref{Fig:SelfConst}(a). If we denote the corrected vertex by $\bar{J}_x$, then
\begin{align}
\bar{J}&_x(\qvec,E) = v_0 I^* \otimes s_x \\ \nonumber
&+ \sum_l w_l \int \frac{d^2 p}{(2\pi)^2} A_l G^R(\pvec,E) \bar{J}_x(\qvec,E) G^A(\pvec+\qvec,E) A_l.
\end{align}
For $\qvec = 0$ we guess a solution of the form $\bar{J}_x(0,E) = fv_0 I^* \otimes s_x$, which gives
\begin{align}
fI^* &\otimes s_x = I^* \otimes s_x + f\sum_l w_l  A_l (I^* \otimes s_x) A_l \\ \nonumber
& \times \int \frac{d^2 p}{(2\pi)^2} \frac{E^2 + \Gamma^2 + v_0^{\phantom{0}2}p_x^2 - v_0^{\phantom{0}2}p_y^2}{[(E+i \Gamma)^2 - v_0^{\phantom{0}2} p^2] [(E-i \Gamma)^2 - v_0^{\phantom{0}2} p^2]}.
\end{align}
Due to $x$-$y$ symmetry the terms with momenta in the numerator vanish. Moreover, $A_l (I^* \otimes s_x) = \xi_l (I^* \otimes s_x) A_l$, where $\xi_l=\pm 1$. Therefore $\sum_l w_l  A_l (I^* \otimes s_x) A_l = (\sum_l \xi_l w_l) (I^* \otimes s_x)$, and the matrix structure of the equation is satisfied. After integrating we find that
\begin{align}
f&= \left( 1 - \half \frac{\sum_l \xi_l w_l}{\sum_l w_l} \right)^{-1},
\end{align}
where we used the fact that $\Gamma \ll E$. We can therefore conclude that the vertex correction is
\begin{align} \label{Eq:beta_ineq}
\frac{2}{3} \leq f\leq 2.
\end{align}

\vspace{5mm}
% ---- Cooperon ----

The next task is to solve the self-consistent equation of the Cooperon, which is depicted in Fig.~\ref{Fig:SelfConst}(b),
\begin{align} \label{Eq:Cooperon}
C&_{(ij)(nm)}(\kvec,\kvec'';E,\Qvec,\omega) = \\ \nonumber
& \int \frac{d^2 k'}{(2\pi)^2}V_{(ij)(i'j')} \Pi_{(i'j')(i''j'')}(\kvec';E,\Qvec,\omega)V_{(i'' j'')(nm)} \\ \nonumber
& + \int \frac{d^2 k'}{(2\pi)^2} V_{(ij)(i'j')} \Pi_{(i'j')(i''j'')}(\kvec';E,\Qvec,\omega) \\ \nonumber
& \qquad\times C_{(i''j'')(nm)}(\kvec',\kvec'';E,\Qvec,\omega), \\
V&_{(ij)(nm)} = \sum_{l} w_l [A_l]_{in} [A_l]_{jm}, \\
\Pi&_{(ij)(nm)}(\kvec';E,\Qvec,\omega) = \\ \nonumber
& \qquad [G^R(\kvec'+\Qvec,E+\omega)]_{in} [G^A(-\kvec',E)]_{jm}.
\end{align}
This equation can be considered as a matrix equation for $C(E,\Qvec,\omega)$, which acts on the vector space $|\kvec\rangle \otimes |ij\rangle$, where $\kvec$ denotes the momenta, and $ij$ denote the internal degrees of freedom of the two particles (of dimension 16).

Anticipating an infrared divergence which is proportional to a diffusive propagator, the Cooperon may be presented as
\begin{align} \label{Eq:Effective_Cooperon}
C_{(ij)(nm)}(\kvec,\kvec'';E,\Qvec,\omega) = & \, c \frac{|d\rangle \langle d |}{D Q^2-i \omega} \\ \nonumber
& + (regular \,\, terms).
\end{align}
Plugging this ansatz into Eq.~(\ref{Eq:Cooperon}), we can extract an equation for the diverging term
\begin{align}
c \left( 1 - V\Pi\right ) \frac{|d\rangle \langle d |}{D Q^2-i \omega} = V\Pi V.
\end{align}
Multiplying from the left with $(V\Pi V)^{-1}$ and from the right with $|d \rangle$, gives an eigenstate equation for the diffusive mode
\begin{align} \label{Eq:Cooperon_eig}
(\Pi V)^{-1} \left(V^{-1} -  \Pi\right) |d\rangle = c^{-1} (D Q^2-i \omega)|d \rangle,
\end{align}
where
\begin{align}
&\left(V^{-1} - \Pi\right) = \left(\sum_l w_l A_l \otimes A_l\right)^{-1} \\ \nonumber
&- \int \frac{d^2 k}{(2\pi)^2} \frac{1}{[(E+\omega+i\Gamma)^2 - v_0^{\phantom{0}2} k^2][(E - i\Gamma)^2 - v_0^{\phantom{0}2} k^2]} \\ \nonumber
& \qquad \times (E+\omega+i\Gamma  + v_0 k_x I^* \otimes s_x + v_0 k_y I\otimes s_y) \\ \nonumber
& \qquad \quad \otimes (E-i\Gamma - v_0 k_x I^* \otimes s_x - v_0 k_y I\otimes s_y).
\end{align}
The terms which are linear in $k_x$ and $k_y$ vanish in the integration, and the remaining three integrals, which multiply the three matrices $(I\otimes I)\otimes(I\otimes I),(I^*\otimes s_x ) \otimes (I^* \otimes s_x)$ and $(I\otimes s_y)\otimes(I\otimes s_y)$, are respectively
\begin{align}
& \int \frac{d^2 k}{(2\pi)^2} \frac{(E+\omega + i \Gamma)(E - i \Gamma)}{[(E+i\Gamma)^2 - v_0^{\phantom{0}2} k^2][(E - i\Gamma)^2 - v_0^{\phantom{0}2} k^2]} \\ \nonumber
& \qquad = -\frac{(\sum_l w_l)^{-1}}{2}, \\
& \int \frac{d^2 k}{(2\pi)^2} \frac{-v_0 ^2 k^2_x}{[(E+i\Gamma)^2 - v_0^{\phantom{0}2} k^2][(E - i\Gamma)^2 - _F^{2} k^2]} \\ \nonumber
& \qquad = \frac{(\sum_l w_l)^{-1}}{4}  + O(\Gamma/E,\alpha), \\
& \int \frac{d^2 k}{(2\pi)^2} \frac{-v_0^{\phantom{0}2} k^2_y}{[(E+i\Gamma)^2 - v_0^{\phantom{0}2} k^2][(E - i\Gamma)^2 - v_0^{\phantom{0}2} k^2]} = \\ \nonumber
& \qquad = \frac{(\sum_l w_l)^{-1}}{4}  + O(\Gamma/E,\alpha).
\end{align}
Therefore
\begin{align} \label{Eq:The_matrix}
&\left(V^{-1} - \Pi\right) \approx \left(\sum_l w_l A_l \otimes A_l\right)^{-1} - \left( \sum_l w_l \right)^{-1} \\ \nonumber
& \times \half \left( 1 - \half \left[ (I^*\otimes s_x)\otimes(I^*\otimes s_x) + (I\otimes s_y)\otimes(I\otimes s_y) \right] \right).
\end{align}

We are interested in the zero mode of the above matrix. Since $V$ mixes all momenta equally, any eigenvector of $V^{-1}$ which depends on momenta will have a diverging eigenvalue, and cannot give rise to a zero mode in the above equation. For generic disorder which respects time-reversal symmetry we find that the zero mode for WTI (with $I^*=I$ for concreteness) and graphene are given by
\begin{align}
\label{Eq:cooperon-vec}
\langle \kvec, ij|d_W \rangle & = \delta_{\tau_i,\tau_j}(\delta_{s_i,1}\delta_{s_j,-1} - \delta_{s_i,-1}\delta_{s_j,1}) /2, \nonumber \\
\langle \kvec, ij|d_{g} \rangle & = \delta_{s_i,s_j}(\delta_{\tau_i,1}\delta_{\tau_j,-1} + \delta_{\tau_i,-1}\delta_{\tau_j,1}) /2, \nonumber
\end{align}
where $\tau_i$ ($s_i$) is the valley (spin/pseudospin) subindex of the index $i$. This can be easily verified by the facts that $A_l |d\rangle = |d\rangle$, $(I^*\otimes s_x)\otimes(I^*\otimes s_x)|d\rangle = -|d\rangle$, and $(I\otimes s_y)\otimes(I\otimes s_y)|d\rangle = -|d\rangle$. Note that the vector $|d\rangle$ has an eigenvalue of 1 with respect to $\Pi V$, and therefore $(\Pi V)^{-1} | d \rangle = |d\rangle$.

The diffusion coefficient $D$ and the constant $c$ from Eq.~(\ref{Eq:Effective_Cooperon}) can be extracted by expanding Eq.~(\ref{Eq:Cooperon_eig}) in $\omega$ and $\Qvec$. After some algebra one finds that in both cases
\begin{align}
c &= 8 v_0^{\phantom{0}2} \frac{\Gamma^2}{E}, \\
D &= \frac{v_0^{\phantom{0}2}}{2\Gamma}.
\end{align}
Note that we keep $c$ although it is of lower order in $\Gamma/E$, since it is associated with the divergence of the Cooperon.

\vspace{5mm}
% ---- Conductivity ----

\begin{figure}[htb]
\begin{center}
\vspace{0cm}
\includegraphics[width=8.5cm,angle=0]{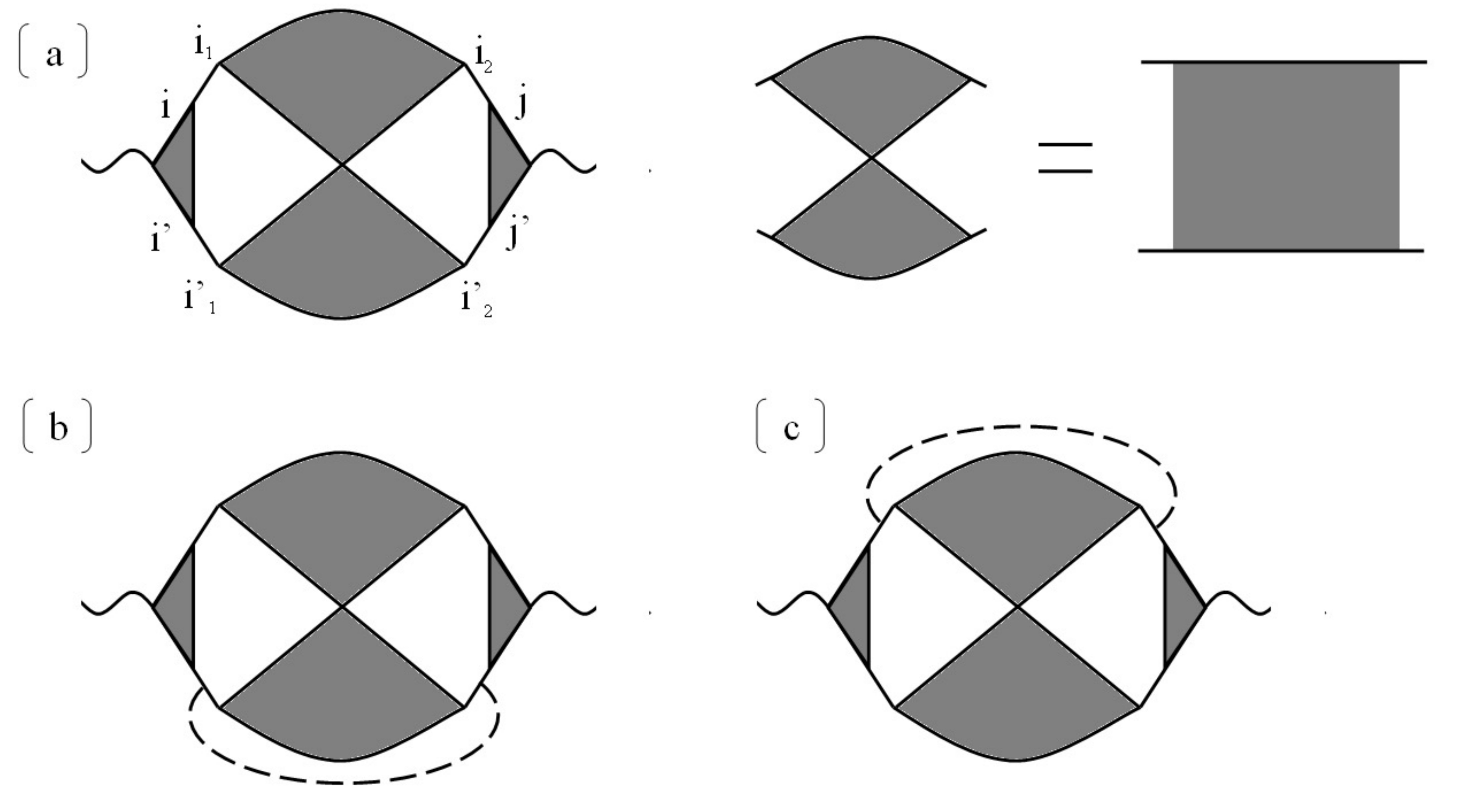}
\vspace{0cm} \caption{ \label{Fig:CondCorrections} %
The leading order quantum corrections to the conductivity. These diagrams can be viewed as a combination of a dressed Hikami box with the Cooperon. }
\end{center}
\end{figure}

The leading term of the conductivity $\sigma_{xx}$ is given by
\begin{align}
\sigma^{0}_{xx} &= \frac{e^2}{2\pi \hbar} \int \frac{d^2 p}{(2\pi)^2} \textrm{Tr} [\tilde{J}_x G^R(\pvec,E_F) J_x G^A(\pvec,E_F)] \nonumber \\
&= \frac{e^2}{\hbar} \frac{E}{\pi v_0^{\phantom{0}2}} D \frac{f_v}{2} = \frac{e^2}{h} \frac{f_v}{2} \frac{E}{\Gamma}.
\end{align}
The first quantum interference correction $\delta \sigma^{a}_{xx} $, which is depicted in Fig.~\ref{Fig:CondCorrections}(a), is given by
\begin{align}
\delta \sigma^{a}_{xx} = \frac{e^2}{2\pi \hbar} & \int \frac{d^2 k d^2 Q}{(2\pi)^4} [\bar{J}_x]_{i' i}[\bar{J}_x]_{j j'} G^A_{j^{'} i^{'}_2}(\kvec,E_F)\\ \nonumber
& \times G^A_{i^{'}_1 i^{'}}(-\kvec+\Qvec,E_F) G^R_{i i_1}(-\kvec+\Qvec,E_F) \\ \nonumber
& \times G^R_{i_2 j}(\kvec,E_F) C_{(i_1 i^{'}_2)(i_2 i^{'}_1)}(\kvec,\kvec;E_F,\Qvec,0).
\end{align}

The divergent contribution to the correction comes from the limit of $\Qvec = 0$ in the Green's functions, where they are regular. Henceforth
\begin{align}
\delta \sigma^{a}_{xx} \approx & \frac{e^2 v_0^{\phantom{0}2}}{2\pi \hbar}f^2\int \frac{d^2 Q}{(2\pi)^2}\frac{c}{DQ^2} \int \frac{d^2 k}{(2\pi)^2} \\ \nonumber
& \times [I^{*}\otimes s_x]_{i' i}[I^*\otimes s_x]_{jj'} G^A_{j' i^{'}_2}(\kvec) G^A_{i^{'}_1 i'}(-\kvec) \\ \nonumber  & \times G^R_{i i_1}(-\kvec) G^R_{i_2 j}(\kvec) \langle i_1 i^{'}_2|d\rangle \langle i_2,i^{'}_1|d \rangle. \\ \nonumber
\end{align}
Keeping only the divergent part of the integral over $\Qvec$, and noticing that the index summation is actually a trace, we have
\begin{align}
\label{Eq:dsig_traceform}
\delta \sigma^{a}_{xx} = & \ln(L) \frac{e^2 v_0^{\phantom{0}2}}{4\pi^2 \hbar}\frac{c}{D} f^2 \int \frac{d^2 k}{(2\pi)^2} \\ \nonumber & \times \textrm{Tr}\left[G^A(-\kvec) (I^* \otimes s_x) G^R(-\kvec) |d\rangle \right. \\ \nonumber
& \qquad \quad \times \left. G^A(\kvec)^T (I^* \otimes s_x)^T G^R(\kvec)^T |d\rangle \right],
\end{align}
where the matrix $|d\rangle$ is defined by $[|d\rangle]_{ij} = \langle ij | d \rangle$. Using the fact that
\begin{align}
&G^A(-\kvec) (I^* \otimes s_x) G^R(-\kvec) \propto 2 v_0^{\phantom{0}2}  k_x k_y (I \otimes s_y) \nonumber \\
& \quad + (E^2 + \Gamma^2 + v_0^{\phantom{0}2}  k_x^2- v_0^{\phantom{0}2} k^2_y)(I^* \otimes s_x) \nonumber \\
& \quad -2E v_0 k_x (I\otimes I) - 2 v_0 k_y \Gamma (I^*\otimes s_z),
\end{align}
the only non vanishing and non negligible traces are those which are proportional to $\textrm{Tr}[(I^* \otimes s_x )|d\rangle (I^* \otimes s_x )|d\rangle]$, $\textrm{Tr}[|d\rangle (I \otimes s_y )|d\rangle (I \otimes s_y)]$, and $\textrm{Tr}[|d\rangle |d\rangle]$. The resulting correction is now
\begin{align}
\label{Eq:dsig_traceform2}
\delta \sigma^{a}_{xx} \approx & -\textrm{Tr}[|d\rangle^2] \ln(L) \frac{e^2 v_0^{\phantom{0}2}}{4\pi^2 \hbar}\frac{c}{D} f^2 \times \nonumber \\
\nonumber & \int \frac{d^2 k}{(2\pi)^2} \frac{E^4 + v_0^{4}k^4 + 4 E^2 v_0^{\phantom{0}2} k_x^2}{[(E+i\Gamma)^2 - v_0^{\phantom{0}2}k^2]^2 [(E-i\Gamma)^2 - v_0^{\phantom{0}2}k^2]^2} \\
\nonumber \approx & -\textrm{Tr}[|d\rangle^2] \ln(L) \frac{e^2 v_0^{\phantom{0}2}}{2\pi^2 \hbar}\frac{c}{D} f^2 \times \nonumber \\
\nonumber & \int \frac{d^2 k}{(2\pi)^2} \frac{4 E^4}{[(E+i\Gamma)^2 - v_0^{\phantom{0}2}k^2]^2 [(E-i\Gamma)^2 - v_0^{\phantom{0}2}k^2]^2}\\
\nonumber = & -\textrm{Tr}[|d\rangle^2] \ln(L) \frac{e^2 v_0^{\phantom{0}2}}{4\pi^2 \hbar}\frac{c}{D} f^2   \frac{E}{16\Gamma^3} \\
= & -T^2 \ln(L) f^2 \frac{1}{4\pi^2} \frac{e^2}{\hbar},
\end{align}
where we replaced $[|d\rangle]^2$ with $T^2$, since they are equal.

As first noted by Ref.~\cite{McCann}, the extra quantum corrections to the conductivity, which are depicted in Figs.~ \ref{Fig:CondCorrections}(b,c), are non vanishing due to the independence on momenta of the current vertex. These two contributions are equal, and are given by
\begin{align}
\delta \sigma^{b}_{xx} = \delta \sigma^{c}_{xx} &= -\frac{f'}{4} \delta \sigma^{a}_{xx},
\end{align}
\begin{align} \label{Eq:f'}
& f' = \frac{1}{4\sum_l w_l} \\
& \qquad \times \left\{ \begin{array}{lc}
            \sum_l w_l \textrm{Tr}[A_l (I\otimes s_z) A^{\phantom{l}T}_l (I\otimes s_z)] & \textrm{WTI} \\
            \sum_l w_l \textrm{Tr}[A_l (\tau_y \otimes s_x) A^{\phantom{l}T}_l (\tau_y \otimes s_x)] & \textrm{graphene}
         \end{array} \right. \nonumber
\end{align}
Since $-1 \leq f' \leq 1$, the sign of the quantum correction is still determined entirely by $T^2$.

We have shown above that $T_g^{\phantom{g}2}=1$ while $T_W^{\phantom{W}2} = -1$. Therefore we can conclude from Eqs.~(\ref{Eq:dsig_traceform2})-(\ref{Eq:f'}) that spinless graphene tends to be localized, while a WTI flows towards perfect conduction.

\section{Weak localization and the time-reversal operator}
\label{App:Trajectories}

In this appendix we provide a straightforward explanation for the fact that the sign of the weak localization correction is the same as the sign of the time-reversal operator squared ($T^2$). To this end, we express the Green's function as a sum over amplitudes associated with trajectories. Similarly, we express the return probability as a sum over products of such amplitudes. The coherent contributions that give rise to weak localization/antilocalization come from products of time-reversal conjugate trajectories. By analyzing the action of $T$ on trajectories the above relation is established.

Consider the Dyson series for the Green's function $G$,
\begin{align}
G = G^0 \sum_{n=0}^\infty (V G^0)^n,
\end{align}
where $G^0$ is the clean Green's function, and $V$ is the disorder potential. The matrix element of $G$ that connects the lattice site $i$ and spin state $\sigma$ with the lattice site $j$ and spin state $\sigma'$ may be written as a sum over trajectories that connect these two sites and spin states, and which go through a series of intermediate points $\alpha = (i\sigma, i_n \sigma_n, i_{n-1}\sigma_{n-1}, \ldots, i_1 \sigma_1, j \sigma')$
\begin{align}
\label{Eq:sum-over}
G_{i\sigma, j\sigma'} &= \sum_{\alpha} \A^{\alpha}_{i\sigma,j\sigma'}, \\
\A^{\alpha}_{i\sigma,j\sigma'} &= G^0_{i\sigma,i_n \sigma_n} \cdot V_{i_{n}\sigma_n, i_{n-1}\sigma_{n-1}} \cdot \ldots \cdot G^0_{i_1\sigma_1,j \sigma'}.
\end{align}

Given that the system is symmetric to some anti-unitary operator, most notably the time-reversal operator $T$, we define $|\bar{\sigma} \rangle = \xi_\sigma T|\sigma \rangle$, where $\xi_\sigma = \pm 1$. Consequently, $G^0_{i\sigma, j\sigma'} = \xi_\sigma \xi_{\sigma'} G^0_{j \bar{\sigma}', i\bar{\sigma}}$ and $V_{i\sigma,j\sigma'} = \xi_{\sigma} \xi_{\sigma'} V_{j\bar{\sigma}',i\bar{\sigma}}$. A straightforward manipulation then yields
\begin{align} \label{Eq:A_bar}
\A^{\alpha}_{i\sigma,i\sigma'} &= \xi_{\sigma} \xi_{\sigma'} \A^{\bar{\alpha}}_{i \bar{\sigma}',i \bar{\sigma}},
\end{align}
where $\bar{\alpha} = (i \bar{\sigma}',i_1 \bar{\sigma_1},...,i\bar{\sigma})$. Note that all the sign factors except $\xi_{\sigma}$ and $\xi_{\sigma'}$ appear twice, and therefore are canceled out.

Using (\ref{Eq:sum-over}) we find that the probability of a particle to return back to its initial site, with perhaps a different spin state, is given by
\begin{align}
\label{Eq:ProductOfPaths}
|G_{i\sigma,i\sigma'}|^2 = \sum_{\alpha,\alpha'} \A_{i\sigma,i\sigma'}^{\alpha} (\A_{i\sigma,i\sigma'}^{\alpha'})^{*}
\end{align}
Two types of pairs of trajectories contribute coherently to the disorder-averaged double sum in equation (\ref{Eq:ProductOfPaths}), since their phases do not fluctuate. The obvious contribution is the classical contribution consisting of pairs with $\alpha = \alpha'$. However, due to $T-$symmetry, an additional contribution exists in which $\alpha$ comes paired with $\bar{\alpha}$. Comparing equation (\ref{Eq:A_bar}) with equation (\ref{Eq:ProductOfPaths}) one finds that pairs of time conjugated paths may appear only if $\sigma' = \bar{\sigma}$. Therefore whenever it appears, the sign factor of such term is $\xi_{\sigma} \xi_{\bar{\sigma}} = \langle \sigma | T^2 | \sigma \rangle = \mathrm{sign}(T^2)$. Also notice that the size of this term is equal to the size of the classical term, and therefore may either double or suppress it. Hence for $T^2 = -1 (1)$ the probability for a diffusing particle to return to its original position is higher (lower) than the classical probability, and this is an indication for weak anti-localization (weak localization). If the Hamiltonian commutes with more than one anti-unitary operator, for example in the case of a spin independent Hamiltonian, the total correction is composed of the contributions from all the different trajectories with $\sigma' = \bar{\sigma}$.

\bibliography{WTI_bib}

\begin{thebibliography}{39}
\bibitem[{Fir()}]{FirstOrder}
\bibinfo{note}{This correction is of zeroth order in $\ln( \Delta/E )/(E\tau)$.
  For a first order corrections see Ref.~\cite{AleinerEfetov}.}
\bibitem[{Flo()}]{Flow_Mong}
\bibinfo{note}{Ref.~\cite{Mong} attempts to find the RG flow of the conductance
  by studying numerically the low energy Hamiltonian, Eq.~(\ref{Eq:Hk_WTI}).
  This study finds that for any disorder strength the flow is towards a perfect
  metal. It is known however, see e.g.~Ref.\cite{AleinerEfetov}, that for
  intermediate and strong disorder in Dirac Hamiltonians, ultra-violet cut-off
  effects become important and may alter the conductance considerably. In
  particular, different cut-offs reflect different topological properties of
  the 3D bulk, most notably the pairs swithcing behavior. We believe that the
  stability of the numerical results to the choice of cut-offs should be
  investigated before the RG flow of the conductance may be concluded from
  numerical calculations.}
\end{thebibliography}

\end{document}